\documentclass[11pt,english]{article}
\pdfoutput=1
\usepackage{jheppub}
\usepackage{amsfonts}
\usepackage{mathtools}
\usepackage{bbm}
\usepackage{multirow}
\usepackage{babel}
\usepackage{verbatim}
\usepackage{mathrsfs}% \usepackage[utf8]{inputenc}
\usepackage{amsmath,amssymb,mathrsfs}
\usepackage{hyperref}

\renewcommand{\d}{\delta}
\newcommand{\lx}{\pounds_X}

\newcommand{\bea}{\begin{eqnarray}}
\newcommand{\eea}{\end{eqnarray}}
\newcommand{\bel}{\begin{align}}
\newcommand{\eel}{\end{align}}
\newcommand{\p}{\partial}
\newcommand{\bes}{\begin{subequations}}
\newcommand{\ees}{\end{subequations}}
\newcommand{\non}{\nonumber}

\title{Dynamical structure of Carrollian Electrodynamics}

\author[a, b]{Rudranil Basu,} \author[b] {Udit Narayan Chowdhury.} 

\author{\\}

\affiliation[a]{Universit\'{e} Libre de Bruxelles and International Solvay Institutes. Campus Plaine C.P. 231, B-1050 Bruxelles, Belgium \\
\& \\
Theoretische Natuurkunde, Vrije Universiteit Brussel, Pleinlaan 2, B-1050 Brussels, Belgium.\\} 
\affiliation[b]{Saha Institute of Nuclear Physics, Block-AF, Sector-1, Salt Lake, Kolkata 700064, India \& \\ 
Homi Bhabha National Institute,
 Training School Complex, Anushakti Nagar, Mumbai 400085, India\\}
\emailAdd{rudranil.basu@ulb.ac.be, udit.chowdhury@saha.ac.in}

\abstract{We present an action of ultra-relativistic electrodynamics on a flat Carroll manifold. The model exhibits a couple of physical degrees of freedom per space-point. We observe that the action of the conformal Carroll algebra on the phase space is Hamiltonian in 4 space-time dimensions. Moreover the elements of the algebra give rise to an infinite number of conserved charges and the charge algebra is an exact realization of the kinematical algebra.}
\begin{document}
\maketitle
\section{Introduction}
One main guiding principle behind building quantum field theory (QFT) models aimed at describing natural particles and interactions is the principle of symmetry. From the standard model of particle physics to the holographic paradigm of AdS/CFT duality, the roles of space-time, internal, and various other accidental symmetries cannot be over-emphasized. For the context of the present article let us focus on the case of conformal field theories (CFT). Unlike in any other space-time dimension, CFTs in 2D enjoy a very large amount of symmetry, in the form of a couple of copies of the infinite dimensional Virasoro algebra. This is the most crucial ingredient behind integrability of 2D CFTs. In addition to this, owing to the infinite symmetry, a plethora of information for such theories including correlation functions can be directly extracted even without any knowledge of the detailed dynamics. In the context of higher dimensional CFTs, there has been a recent upsurge in interests, in exploiting generic symmetry structures much under the umbrella of the conformal bootstrap program, which have met with enormous success \cite{Simmons-Duffin:2016gjk}.

Larger symmetry groups in a system obviously mean stronger analytical control over its predictability. In this view, the pursuit for physical theories with infinite amount of symmetries, like that in 2D CFTs, is an extremely lucrative one. Such a possibility came forward with the advent of infinitely extended Schrodinger algebra studied extensively by Henkel et al (see \cite{Henkel:2002vd} and references therein) and Galilean conformal algebra (GCA) \cite{Bagchi:2009my}, \cite{Bagchi:2009ca}. The later, more suited for motivational purpose of the present work, describes space-time symmetry for non-relativistic conformal systems in arbitrary space-time dimensions. Later studies revealed GCA to be a special case in a family of Lie algebra of diffeomorphisms which are symmetries of non-Riemannian space-time manifolds \cite{Duval:2009vt}, just as the Poincare algebra is a symmetry of Minkowski space-time. These are named Newton-Cartan (NC) manifolds.

In terms of geometric structures, NC manifolds are quotients of Bargmann manifolds. These are 1 higher dimensional Riemann manifolds (with Lorentz signature), equipped with a special globally defined null, parallel transported vector field. For example if $g$ is the metric tensor and $\xi$ is the null vector field in a Bargmann manifold $\mathcal{B}$ then the codimension-1 quotient manifold $\mathcal{B}/\star$ is NC. Here $\star$ is the one-dimensional group of diffeomorphisms generated by $\xi$. Interestingly there is another independent `dimensional reduction' of Bargmann space-time. To see this, first associate the unique 1-form $\theta = g(\xi)$ dual to the vector field $\xi$. Since $d \theta =0 $ due to the parallel transportation property of $\xi$, the distribution defined by the vector fields in the kernel of $\theta$ is integrable, and hence defines a codimension-1 foliation of $\mathcal{B}$. Each of these foliations is a Carroll manifold \cite{Duval:2014lpa}, the intrinsic definition of which will be reviewed in the main body of the paper. From the point of view of Bargmann space-time therefore, NC manifolds and Carroll ones are dual to each other, in terms of geometric structures.

Our interest in Carroll manifolds stems mainly from the aspect of infinite dimensional symmetry property including conformal symmetry. The key point in this direction is that independent vector fields that preserve the structures of a Carroll manifold, even upto a conformal factor, form an infinite dimensional Lie algebra, similar to the case of GCA for a NC manifold. When equipped with conformal structure, this infinite dimensional algebra is named conformal Carroll algebra (CCA). One common feature of CCA for Carroll manifolds of diverse dimension is its infinite dimensional Abelian ideal, formed by arbitrary space-dependent temporal translations named as super-translations. Intrigued by this infinite dimensional space-time symmetry, an ambitious program was started to look for physical systems on these backgrounds which may possess these symmetries. For the case of GCA, the initial observations were regarding establishing gauge theory models with GCA as symmetry \cite{Bagchi:2014ysa, Bagchi:2015qcw, Bagchi:2017yvj}. Vanishing of mass is tied with conformal invariance in certain dimensions and those theories of vectors in Minkowski space enjoy gauge invariance for the same reason. Therefore it is not of surprise that the above mentioned Galilean (or non-relativistic) theories enjoy conformal invariance. In a parallel, studies in ultra-relativistic (UR) models of conformally invariant gauge theories were carried out, which possesses infinite conformal CCA as symmetry \cite{Bagchi:2016bcd}. The present article is a part of the continuum of the later program. 

These observations in \cite{Bagchi:2014ysa, Bagchi:2015qcw, Bagchi:2017yvj, Bagchi:2016bcd} were made at the equations of motion. However for deeper understanding in classical dynamics and to build up a quantization program an action principle is called for. Even without an action, if one is able to append a phase space structure with the equations of motion, a large amount of information can be extracted. Led by these curiosities, action principle was constructed recently \cite{Bergshoeff:2015sic} for the Galilean theories of Electrodynamics, the first example of a gauge theory. However it was noticed later that the constructed action \cite{Festuccia:2016caf}, which contains an additional scalar field on top the theory studied in \cite{Bagchi:2014ysa}, does not enjoy the infinite invariance of GCA. More strikingly that theory lacks any local degrees of freedom \cite{rb}. The present work revolves around a Carrollian version of Electrodynamics and we here propose an action for the same. %Satisfactorily the theory is invariant under infinite dimensional CCA in 3+1 dimensions. We successfully construct the corresponding infinite number of independent conserved charges which commute among themselves.

Before delving directly into the subject matter and results of this article, let us briefly mention the relevance and importance of Carrollian physics in diverse contexts in light of recent literature. Most important of these is the connection with asymptotic symmetries of flat-space time. It was discovered long time back that asymptotic symmetries of 4D asymptotically flat-space times, which are solutions of Einstein equation form an infinite dimensional group, named BMS \cite{Bondi:1962px, Sachs:1962zza}, of which Poincare is a subgroup. The Lie algebra of BMS group is same as that of CCA for a Carroll manifold in 3 dimensions (\cite{Duval:2014uva} and references therein). From this connection with BMS, the Abelian ideals of CCA are named super-translations. The same correspondence holds in one lower dimension as well. The asymptotic symmetry of 3 dimensional asymptotically flat gravitating system, known as BMS${}_3$ \cite{Barnich:2006av} is isomorphic to GCA in 2 dimensions \cite{Bagchi:2010eg}. Again in 2 dimensions, GCA and CCA are isomorphic. These results boosted investigations in the much sought after flat-holography program \cite{Bagchi:2013hja, Bagchi:2012xr} as a holographic principle complementary to the standard AdS/CFT. Apart from the context of flat-holography in 3 dimensions, this Carroll algebra or BMS${}_3$ as well as its supersymmetric extension was shown to be residual worldsheet symmetry of Bosonic \cite{Bagchi:2015nca} and super-string theory \cite{Casali:2016atr, Bagchi:2016yyf} at the tensionless limit respectively. Relevant to our work of constructing explicit models, there has been successes in 2 dimensions, where BMS${}_3$ invariant interacting theories have been constructed \cite{Barnich:2013yka, Gonzalez:2014tba}. However, arguably the most important recent advances in BMS physics, led by Strominger is establishing BMS as symmetry of quantum gravity (at the infra-red) S-matrix and relating this with Weinberg's soft graviton theorem as a consequence of Ward identity corresponding to this symmetry. See \cite{Strominger:2017zoo} for a comprehensive review and references. There is recent excitement in the sector of developing Caroll gravity, by gauging an ultra-relativistic contraction of Poincare algebra \cite{Hartong:2015xda, Bergshoeff:2017btm}. The resultant first order theory describes a dynamical theory of geometrical structures of Carroll manifold. 

It has already been mentioned that the present work is aims at advancing the active program of finding higher dimensional field theory models with infinite symmetry. The model in context is an ultra-relativistic limit of free Electrodynamics. It was noted earlier that, as with the case of Galilean limit \cite{LBLL}, there are two distinct ultra-relativistic limits of Electrodynamics depending upon how the Carrollian limits are taken on the temporal and the spatial components the original Lorentz covariant 4-vector potential. These are named respectively the electric and the magnetic limits. In this article we successfully construct an action for the electric sector of Carrollian electrodynamics. Performing the canonical analysis, it is revealed that:
\begin{itemize}
\item the theory possessed 2 propagating degrees of freedom per space-point, like in Maxwell electrodynamics and
\item gauge invariance is still governed by the $U(1)$ gauge group.
\end{itemize}
The most significant part of the analysis involves the CCA symmetry of the action. Avoiding the standard Noetherian route of finding conserved quantities corresponding to symmetry generators of CCA, we take the pre-symplectic \cite{Crnkovic:1986ex} point of view. We show that the generators of CCA indeed generate Hamiltonian flows on the phase space of solutions, proving CCA to be symmetry of the phase space. Claiming the conformal generators to be Hamiltonian forces the space-time dimension to be 4 and fixes the classical conformal dimensions of the fields, as in Maxwell theory. Then we check that corresponding Hamilton functions are gauge invariant and time-preserved, therefore qualifying as conserved charges. We then establish a linear homomorphism between an algebra (induced by the pre-symplectic structure) of those charges and CCA. This is a statement of exact realization the Carrollian symmetry algebra at the level of charge algebra. No central extensions are encountered for the infinite dimensional ideal subalgebra of CCA.

In section 2, we review geometric properties of Carroll manifold and emergence of CCA as algebra of vector fields whose diffeomorphisms preserve the Carrollian structure upto a conformal factor. We then focus on CCA for `flat-Carroll' manifolds and review the highest weight representation suitable for the fields in Carrollian electrodynamics. We put all the dynamical content in section 3, including the introduction of an action, the canonical analysis and the pre-symplectic analysis leading to the charge algebra. We conclude with a few future directions of the present work. In the appendix, in order to draw a parallel for Maxwell electrodynamics we perform a similar covariant phase space framework and show how the kinematical algebra of background space-time's Killing vector fields are realized at the level of charges.
\section{The Conformal Carroll Algebra}
In order to make a self-consistent analysis, we would briefly review the symmetries and structures of Carrollian manifolds following \cite{Bagchi:2016bcd, Duval:2014lpa}. As an invitation to it, we here take the approach of viewing Carrollian symmetries descending from Poincare symmetries of Minkowski manifold at an ultra-relativistic (UR) limit and a natural emergent infinite dimensional extension of the symmetry algebra. This point of view is best suited for our purpose of building a field theory on a Carroll space-time based on representation theory. Of course we would supplement this discussion with a description of the general geometric structures of a Carroll space-time later and review the symmetry algebra in self-consistent manner. Later in this section we will describe a representation of the symmetry algebra suited to describe UR field theories.
\subsection{Conformal Carroll isometries}
An ultra-relativistic Inonu-Wigner contraction of the Poincare algebra for $d+1$ space-time dimensions leaves us with the finite part of the Carrollian algebra with following non-vanishing Lie-brackets:
\bea \label{ca-finite} 
&& [J_{ij},B_{k}]=\delta_{k[j}B_{i]}, \hspace{.2cm}[J_{ij},P_{k}]= \delta_{k[j}P_{i]},\hspace{.2cm}   
[B_{i},P_{j}]=-\delta_{ij}H \non\\
&& [J_{ij},J_{kl}] = J_{i[l} \d_{k]j}+J_{j[k} \d_{l]i}
%,\hspace{.2cm} [D,K]=-K, \non\\ &&\hspace*{.2cm} [K,P_{i}]=-2B_{i}, \hspace*{.2cm}[K_{i},P_{j}]=-2D\delta_{ij}-2J_{ij}, \hspace*{.2cm} [H,K_{i}]=-2B_{i}\\
%&&[D,H]=H, \hspace*{.2cm}[D,P_{i}]=P_{i}, \hspace*{.2cm} [D,K_{i}]=-K_{i} \non.
\eea
where $J\in \mathfrak{so}(d), B, P$ and $H$ are to be respectively interpreted as generators of spatial rotations, Carrollian boost, the spatial translation and time translation. Moreover a generalization to finite conformal Carroll algebra can be obtained via the same contraction if we consider generators of conformal isometries of Minkowski space. We include the new generators as Dilatation ($D$), temporal and spatial special conformal generators, ($K, K_i$) which are $\mathfrak{so}(d)$ scalar, scalar and vector. Apart from the usual $\mathfrak{so}(d)$ transformation rules, non-trivial Lie-brackets of these new generators with those of \eqref{ca-finite} are
\begin{align} \label{cca-finite}
& [K,P_{i}]=-2B_{i}, \hspace*{.2cm}[K_{i},P_{j}]=-2D\delta_{ij}-2J_{ij},\, [K_i, B_j] = -\d_{ij}K \non \\
& [H,K_{i}]= 2B_{i},\, [D,H]=-H,\, [D,P_i] = -P_i,\, [D,K_{i}]=K_{i}
\end{align}
The above generators can also be viewed in terms of vector fields on a Carroll manifold. The following special limit:
\bea \label{sptm_scaling}
x^{i} \rightarrow x^{i},~~ t \rightarrow \varepsilon t, \quad \varepsilon \to 0
\eea
on the Minkowski isometry vector fields in a regularized way, does the job of implementing the UR limit. For example, the Carrollian boost is obtained from that on Minkowski space as:
\bea
B^{(\mathrm{Minkowski})}_i = t \p_i + x^i \p_t \rightarrow \varepsilon t \p_i + \frac{1}{\varepsilon}x^i \p_t \xRightarrow[]{\text{regularize}} B^{(\mathrm{Carroll})}_i = x^i \p_t
\eea
The list of all the generators appearing in \eqref{ca-finite} and \eqref{cca-finite} in terms of vector fields now reads
\bea \label{vef}
&& J_{ij}= (x_{i}\partial_{j}-x_{j}\partial_{i}),\hspace{.2cm} D=(t\partial_{t}+x^{i}\partial_{i}),\hspace{.2cm}K= (x^{k}x_{k})\partial_{t} \non \\ 
&& B_{i}= x_{i}\partial_{t}, \hspace{.2cm}H=\partial_{t},\hspace{.2cm} P_{i}=\partial_{i},\hspace{.2cm}
K_{i}= 2x_{i}(t\partial_{t}+x^{k}\partial_{k})-(x^{k}x_{k})\partial_{i}.
\eea
It is easy to notice that the subset of generators: $\{J_{ij},P_i, D, K_i\}$ form the conformal algebra of $d$ dimensional Euclidean space. On the other hand, we can append to this set an arbitrarily large number of generators, all captured in the form:
\bea \label{sutra}
M_f = f(x) \p_t
\eea 
which still closes in terms of the Lie-bracket algebra with the rest. Here $f$ are arbitrary tensors expressed as functions of the original spatial coordinates ($x^i, i =1, \dots ,d$) transforming in irreps of $\mathfrak{so}(d)$. For example the choices $f =1, x^i, x^k x_k$ respectively give $H, B_i, K$. $M_f$ together with $\{J_{ij},P_i, D, K_i\}$ give the infinite dimensional conformal Carrollian algebra (CCA). The rest of the Lie brackets involving the finite set $\{J_{ij},P_i, D, K_i\}$ and the infinite set $\{M_f\}$ are:
\bes \label{cca}
\bea
&& [J_{ij},M_f] = M_g, \quad g = x_{[i} \p_{j]} f \\
&& [P_i, M_f] = M_{\p_i f} \\
&& [D, M_f] = M_h, \quad h = x_i \p_i f - f \\
&& [K_i, M_f] = M_{\tilde{h}} \quad \tilde{h} = 2x_i h - x^k x_k \p_i f. 
\eea
\ees 
The set of $M_f$ for all $f$ form an Abelian ideal of CCA.

Specializing to the case of $d=2$ and choosing the topology of the background to be $\mathbb{R} \times S^2$ instead of $\mathbb{R}^3$, this algebra becomes isomorphic to the BMS$_4$ algebra \cite{Sachs:1962zza} \cite{Bondi:1962px}. This generates the asymptotic symmetry group of 4 dimensional asymptotically flat space-time at null infinity. The finite part $\{J_{ij},P_i, D, K_i\}$ in that case corresponds to the base sphere's conformal isometry and the arbitrary space dependent time translations are named as super-translations. The role of the time coordinate in that case, is taken by the retarded null-time.

The CCA generating vector fields, which were shown to descend from Minkowski structures, also emerge from an elegant analysis intrinsic to Carroll manifolds, without reference to Minkowski or any other Riemann manifold. In this approach, a Carroll manifold \cite{Duval:2014lpa} is a differentiable manifold equipped with a rank-2 symmetric covariant tensor $g$ and a nowhere vanishing vector field $\xi$, such that $i_{\xi} g = 0$ at all points. For our purpose, we need not require a connection for now.

The obvious way to define conformal Carroll isometries are via the diffeomorphisms which preserve the above structures, upto conformal factors. For example, if the vector field $X$ generates such a diffeomorphism, then it should satisfy:
\bea \label{isomet}
\pounds_X g = \kappa\, g, \, \, \pounds_X \xi = \lambda\, \xi ,
\eea
where for our purpose we will chose $\kappa, \lambda$ to be constants satisfying $ \kappa + 2 \lambda = 0$.

Then for the choice of `flat-Carroll' manifold, we have:
\begin{align} \label{fc}
& g = \d_{ij} dx^i \otimes dx^j, \, \, \xi = \p_t,  \\
&\mbox{hence manifestly }~~  i_{\xi} g =0 \non .
\end{align}
Therefore we can solve directly for $X$ in the differential equations \eqref{isomet} to get most general solution:
\begin{align} \label{gen_iso}
X = p^i P_i + \omega^{ij} J_{ij} + \Delta D + k_i K^i + M_f
\end{align}
where $p^i, \omega^{ij}, \Delta, k_i$ are integration constants and the basis vector fields are same as those appearing in \eqref{vef} and \eqref{sutra}.
\subsection{Representation of CCA}
The goal of the present article is to describe understand symmetries of an UR field theory, ie that on a flat-Carroll manifold equipped with \eqref{fc}. A natural geometrical strategy of approaching this is by considering the classical fields to be tensors on the manifolds, on which diffeomorphisms like \eqref{gen_iso} act by Lie derivative, particularly if we don't deal with Fermions. On the other hand, from a particle physics point of view, it would be natural to describe fields in terms of their spins \footnote{For Minkowski field theories, both point of views are equivalent since definite integral spin (helicity) representations of the Poincare algebra are equivalent to tensors on the space-time.}. We here take the second, traditional path, followed in our earlier works \cite{Bagchi:2015qcw, Bagchi:2016bcd}.

We should have the fields to transform in irreps of the rotation generating subalgebra $\mathfrak{so}(d)$ and to have definite scaling dimension under the generator $D$, taking their commutativity $[J_{ij}, D]=0$ into our favor. For an example, if one specifies to $d=3$, this means that we will label our fields with integral spins and conformal weights; precisely at space-time origin:
\begin{align}
\d_{J^2} \Phi (0,0) = L(L+1) \Phi (0,0) , \qquad \d_D \Phi (0,0) = \Delta \Phi(0,0),
\end{align}
supplemented with:
\begin{align}
\d_{H}\Phi (x,t) = \p_t \Phi (x,t), \quad \d_{P_i}\Phi (x,t) = \p_i \Phi (x,t), \quad \d_{K_i} \Phi (0,0)=0.
\end{align}
For the `super-translation' generators $M_f = f(x) \p_t$ we would, for now choose $f$ to be homogeneous polynomials of the spatial coordinates so that they fall into irreps of the rotation group and have definite scaling dimension. In particular if $f$ is a degree $\alpha$ polynomial, then $[D, M_f] = (\alpha -1)M_f$. As a statement of highest weight representation \footnote{We hereby note that, as described in the introduction, CCA for Carroll manifolds of dimensions 2 and 3 respectively are equivalent to asymptotic symmetries of asymptotically flat gravitating systems of 3 and 4 dimensions, respectively and these symmetries go in the name of BMS group. Unitary induced representations, which are different from the present one are discussed recently in \cite{Barnich:2014kra, Banerjee:2018gce}}, we further impose:
\bea
\d_{M_f} \Phi (0,0) = 0, \quad \mbox{if} \, \, \alpha > 1
\eea
The last piece required in fully specifying the representation of CCA in terms of fields is the action of the boost operator $B_i$ which corresponds to $M_f = x^i \p_t$ and obviously is the case of $\alpha =1$. At this stage, we don't fully fix this and keeping in mind that we wish to focus only for $\mathfrak{so}(d)$ scalars and vectors, we keep this indeterminacy upto a couple of constants:
\begin{align}
\d_{B_i} \Phi^{L=0}(0,0) = \kappa_1 \Phi^{L=1}_i(0,0) \, , \, \, \d_{B_i} \Phi^{L=1}_j(0,0) = \kappa_2 \delta_{ij}\, \Phi^{L=0}(0,0) 
\end{align}
From now on we won't explicitly write spin of the operator; the index structure will be sufficient to clarify this.

Equipped with this, we can now write the action of the CCA generators on the $\mathfrak{so}(d)$ scalar and vector fields having definite scaling dimension as (for details of the derivation, we refer to our earlier work \cite{Bagchi:2016bcd}):
\begin{subequations} \label{symm_action}
\bea\label{boost-sc-vect} 
%\d_{B_{i}}\Phi (x,t) &=& x_{i}\partial_{t}\phi+ \kappa_1 \Phi_{i},\quad \d_{B_{i}}\Phi_{j} (x,t) = x_{i}\partial_{t}\Phi_{j} + \kappa_2\, \delta_{ij}\Phi \\
\mathrm{Translation:} ~&& \d_{p} \Phi (x,t) = p^j \p_j \Phi, ~~ \d_{p} \Phi_i (x,t) = p^j \p_j \Phi_i \\
\mathrm{Rotation:}~&&\d_{\omega}\Phi (x,t) = \omega^{ij}(x_i \p_j- x_j \p_i) \Phi ,\\ 
                   && \d_{\omega}\Phi_l (x,t) = \omega^{ij}\left[(x_i \p_j- x_j \p_i) \Phi_l + \d_{l[i} \Phi_{j]}\right], \\
\mathrm{Dilatation:}~&& \d_{\Delta} \Phi (x,t) =  (t \partial _t +x^i \partial _i + \Delta) \Phi, \\
                     && \d_{\Delta} \Phi _j(x,t) =  (t \partial _t +x^i \partial _i + \Delta) \Phi_j,	\\
\label{spc1}\mathrm{Sp. conformal}~&& \d_{k} \Phi (x,t)= 2 k^i\left[(\Delta x_{i}+ x_{i}t\partial_{t}+ x_{i}x^{j}\partial_{j} - \frac{x^{j}x_{j}}{2} \partial_{i})\Phi +\kappa_1\, t\Phi_{i} \right], \\
\label{spc2} && \d_{k}\Phi_{l}(x,t)= 2 k^i(\Delta x_{i} +x_{i}t\partial_{t}+x_{i}x^{j}\partial_{j}- \frac{x^{j}x_{j}}{2} \partial_{i})\Phi_{l} +2 k_l x^{j}\Phi_{j} \non\\ 
              && ~~~~~~~ \qquad \quad -2 k^i x_{l}\Phi_{i} + 2\kappa_2 k_l \, t \Phi , \\
  \label{M_on_scalar} \mathrm{Supertranslation} && \d_{M_f}\Phi(x,t)= f(x) \p_{t}\Phi +\kappa_1 \Phi_i \p_i f(x),\\
 \label{M_on_vector}&& \d_{M_f}\Phi_{i}(x,t)]= f(x)\p_{t}\Phi_{i} + \kappa_2 \Phi \, \p_i f(x).
\eea
Here we have explicitly used the parameters of transformations, constants $p^i$ for spatial translation, antisymmetric $\omega^{ij}$ for spatial rotation, $\Delta$ for scaling, $k^i$ for spatial part of special conformal transformation and the arbitrary function $f(x)$ for super-translations including time translation (for $f=1$), Carrollian boost (for $f = x^i$) and temporal part of special conformal transformation (for $f = x^2$), as they appear as coefficient of the basis vector fields in \eqref{gen_iso}.

Note that scaling dimensions for both the spin-0 and spin-1 fields must be same in order to hold the homomorphism $\left([D, B]  \rightarrow [\d_D, \d_{B_i}] \Phi \right)$ true.
\end{subequations}
\section{Carrollian Electrodynamics}
As a first dynamically non-trivial example of a field theory on a Carroll space-time, we studied \cite{Bagchi:2016bcd} electrodynamics in the UR limit. The basic assumption \footnote{This is plausible because that stems from solution of the equation $dF =0$ for a field strength 2-form $F$. Since this statement is independent of any other structure on a differential manifold, the potential formulation should stand for Riemannian or Carrollian manifolds alike.} of this line of study was the existence of a potential formulation of UR electrodynamics. In addition to this, we stick to a potential based formulation since this approach can be generalized to theories with interactions, like to gauge fields coupled to matter and/ or where non-Abelian gauge invariance is present. This is in contrast to the field strength based exposition in \cite{Duval:2014uoa}.

Motivated by an approach for Galilean electrodynamics \cite{Bagchi:2014ysa}, we derived equations of motion for UR field theories in \cite{Bagchi:2016bcd}. This essentially requires scaling the temporal and spatial components of four-vector potential field of relativistic Maxwell theory in addition to the space-time coordinate scaling \eqref{sptm_scaling} required for going to Carrollian regime. As it turns out \cite{Duval:2014uoa}, there exists a couple of independent such scalings which give rise to dynamically interesting equations of motion for UR electrodynamics. The scaling rules:
\bes \label{field_scaling}
\bea
\label{elec_sc} && A_t \rightarrow A_t, \, A_i \rightarrow \varepsilon A_i \quad \mbox{(electric sector)} \\
\label{mag_sc} && A_t \rightarrow \varepsilon A_t, \, A_i \rightarrow  A_i \quad \mbox{(magnetic sector)}
\eea
\ees
which are respectively called the electric and the magnetic sectors respectively. The naming of the sectors will become apparent shortly breaks. Note that the different way of scaling the temporal and the spatial parts break Minkowski boost, as expected.

The same set of scaling rules \eqref{sptm_scaling} and \eqref{field_scaling} simultaneously can be applied on the relativistic equation of motion $d \star F = 0$ (turning off the source) to derive  equations of motion of UR electrodynamics. From now on, we will denote the $\mathfrak{so}(d)$ scalar potential $A_t$ in the UR regime as $B$. Respectively the electric and magnetic sector equations of motion are:
\bes
\bea
\label{eleom} && \p^{i}\p_{i}B-\p^{i}\p_{t}A_{i}=0,~~ \p_{t}\p_{i}B-\p_{t}\p_{t}A_{i}=0; \\
\label{mageom} && \p^{i}\p_{t}A_{i}=0,~~ \p_{t}\p_{t}A_{i}=0.
\eea
\ees
The spatial $\mathbb{R}^d$ leaves on a flat-Carroll manifold \eqref{fc}, get an induced flat non-degenerate metric $\delta_{ij}$. This means that from now on, we need not distinguish between covariant or contravariant $\mathfrak{so}(d)$ tensors and parallel to these. 

Note that in the Magnetic sector \eqref{mageom}, the scalar potential does not occur. Also, in the electric sector if we define the Carrollian electric field $E_i = \p_t A_i - \p_i B$ in an analogous way to the relativistic case, the equations of motion \eqref{eleom} read:
\begin{align}
\p_i E_i =0, \quad \p_t E_i =0
\end{align}
exactly as appearing in \cite{Duval:2014uoa}, which reflects the disappearance of the Ampere current term involving the magnetic field.

We should also keep in mind that the present algebraic formulation of a physical theory in terms of $\mathfrak{so}(d)$ scalars and vectors does not manifestly reflect covariance with respect to Carrollian geometry. But one of the successes of this approach was that the equations of motion were shown to invariant under CCA \cite{Bagchi:2016bcd}, whose generators act on the fields as in \eqref{symm_action} with specified values for the constants $\kappa_1 =0, \kappa_2 =1$ for the electric sector and $\kappa_1 =1, \kappa_2 =0$ for the magnetic one. On the other hand, the price for this is that a generalization to non-flat Carrollian manifold is not straightforward. 
\subsection{The Lagrangian}
While trying to figure out whether the above equations motion can be derived from an action, it immediately comes into notice that the magnetic ones \eqref{mageom} themselves cannot come from one. This is because there are a couple of equations involving the same variable $A_i$. Therefore we defer the discussion for the Magnetic case to a later investigation and focus to the electric sector in the present analysis.

Let us start by proposing a Lagrangian:
\begin{align}
\label{carlag}
L=\int_{\Sigma} d^d x \, \left( \dot{A_i}^2 + (\p_i B)^2 - 2  \dot{B} \p \cdot A \right)
\end{align}
where we have adopted short hand notations $\p_t B = \dot{B}, \p \cdot A = \p_i A_i$ etc and $\Sigma$ is a spatial leaf with topology of $\mathbb{R}^d$ defined by $t=$constant such that the vector field $\p_t$ is a degenerate direction of the flat-Carroll bilinear form, as in \eqref{fc}. For simplicity, as we specialize to the flat-Carroll case \eqref{fc}, $\Sigma$ has the flat metric $\d_{ij}$ of $\mathbb{R}^d$. The equations of motion coming out of the action $S=\int dt L$ are identical to those found from the limiting approach \eqref{eleom}.
%\bes \label{careom}
%\begin{align}
%\label{car1}
%& \ddot{A}_i-\partial_{i}\dot{B}=0\\
%\label{car2}
%& \p^2 B - \p \cdot \dot{A}=0
%\end{align}
%\ees
In dealing with the variational principle we have chosen boundary conditions on fields and allowed variations such that the terms:
\bea
\p_i B \d B \Big{|}_{\p \Sigma} = 0 = \dot{B} \d A_i \Big{|}_{\p \Sigma}.
\eea
% dropped terms at spatial infinity for the variational principle to work.These are the equations of motion for Carrollian electrodynamics in the Electric sector. 
%Since this action formulation is manifestly not defined in terms of a Lagrangin, which is a density in the Carrollian sense, 
We postpone the discussion of its Carrollian invariance, which is not manifest from the action, to a later section. Rather, what more explicit is its gauge invariance. The above action is invariant under the following gauge transformations:
\begin{align} \label{cargt1}
& \d_{\alpha} B = \dot{\alpha}_1, \, \d_{\alpha} A_i = \p_i \alpha_2 \non\\
& \mbox{if }\, \alpha_1 - \alpha_2 = \mbox{space-time constant}
\end{align}
Before going on to an analysis of the global Carrollian symmetries and the dynamical realization of CCA, we perform a Dirac constraint analysis of the system, since this is obviously plagued with gauge redundancy \eqref{cargt1}.

%\textit{Include the case of Magnetic sector here.}
\subsection{Canonical analysis}
While performing Legendre transformation on \eqref{carlag} to go over to a Hamiltonian analysis, we encounter a primary constraint:
\begin{align} \label{constr1}
C_1 = 2 \p \cdot A + \pi_B \approx 0
\end{align}
However time derivative of the $A_i$ field is solved via invertible momentum relation:
\begin{align}
\pi_A^i = 2 \dot{A}_i.
\end{align}
We therefore augment the canonical Hamiltonian with a Lagrange multiplier:
\begin{align} \label{car_Ham}
H = \int_{\Sigma} d^d x \left(\frac{1}{4} (\pi^i_A)^2 - (\p_i B)^2 + u_1 (2 \p \cdot A + \pi_B) \right)
\end{align}
Time preservation of the primary constraint \eqref{constr1} now gives us a secondary one:
\begin{align}
\{ C_1, H\} = \p \cdot \pi_A -2 \p^2 B = C_2  \approx 0
\end{align}
Time preservation of $C_2$ neither gives rise to a new constraint nor does it solve for the Lagrange multiplier. Hence $C_1, C_2$ are the only constraints. We can smear them with test functions to define
\begin{align}
\mathcal{C}_1[\lambda] = \int d^d x \, \lambda \left( 2 \p \cdot A + \pi_B\right), \, \mathcal{C}_2[\lambda] = \int d^d x \lambda \, \left( \p \cdot \pi_A -2 \p^2 B\right)
\end{align}
and observe that they Poisson commute:
\bea
\{\mathcal{C}_1[\lambda_1],\mathcal{C}_2[\lambda_2] \} =0
\eea
making both the constraints first class.

In order to see the gauge transformation generated by them, let us define a gauge generator made out of these constraints:
\begin{align}
G = \mathcal{C}_1[\lambda_1] + \mathcal{C}_2[\lambda_2].
\end{align}
The dynamical variables transform under gauge transformation:
\begin{align} \label{cargt2}
\d_G B = - \lambda_1, \, \, \d_G A_i = \p_i \lambda_2.
\end{align}
But these are not independent as their gauge transformation and time evolution must commute \cite{Banerjee:1999hu, Banerjee:1999yc}, giving rise to the relations:
\begin{align} \label{cargtrel}
\p_i (\lambda_1 + \dot{\lambda}_2) = 0.
\end{align}
This should be compared with the gauge freedom observed at the level of equations of motion in our earlier work \cite{Bagchi:2016bcd}. The statements of gauge invariance \eqref{cargt2} and \eqref{cargtrel} derived dynamically via constraint analysis are equivalent to the ones \eqref{cargt1} stated at the level of action through inspection.

Since there are only two scalar first class constraints, the physical phase space dimension is $2\times (d+1)-2 \times 2 =2d-2$ per space point, leaving $d-1$ degrees of freedom per space point just like Minkowski electrodynamics.
\subsection{Pre-symplectic analysis and Carrollian symmetries of UR symmetries}
Minkowski field theory actions are constructed to be manifestly Poincare invariant. More generally, field theories on arbitrary Riemann manifolds are constructed covariantly and the isometries (if there are any) of the manifold are automatically global symmetries of the action. However, the particular action for UR electrodynamics that comes from the Lagrangian \eqref{carlag} does not manifestly reflect the Carrollian symmetries \eqref{symm_action}.

The main goal of the present article is to see if the conformal Carrollian symmetry algebra is realized at the level of charges in the context of Carrollian electrodynamics described above. This of course involves first checking whether these Carrollian `isometries' are at all symmetries of the theory and then finding corresponding conserved charges. The second level is to see the charge algebra realization of the symmetry algebra. Standard Noether method progresses by finding the charges, if the symmetry transformation leave the action invariant. One then goes on calculating Poisson (or Dirac if second class constraints are involved) brackets of the charges expressing them in terms of canonical variables.
 
We here proceed on an alternative yet equivalent route. At the initial stage, instead of checking whether the Carrollian symmetries are actually symmetries of the action, we prefer to perform a milder check. That is checking the symmetries of the phase space structure or rather the pre-symplectic \footnote{The name pre-symplectic is necessary instead of symplectic because the action has gauge invariance as understood from the above analysis.} structure on the space of solutions that arises from the action. From now on, we will call the space of solutions equipped with a pre-symplectic structure, the phase space, which is different from the canonical phase space. This analysis will serve a couple of purposes at a single go. First of all, the transformations which preserve the pre-symplectic structure, give rise to locally Hamilton functions. Those Hamilton functions which are time preserved, qualify as conserved charges. Secondly, the algebra of charges is more easily read out via the moment map derived from the pre-symplectic structure, rather than computing the Poisson (or Dirac) brackets of standard Noether charges. Moreover the homomorphism between the Lie-algebra of Hamilton vector fields and the algebra of charges is guaranteed in this formalism, modulo central elements. As a side-note, we keep in mind that the symmetry group generated by conserved charges is in general a subgroup of that generated by Hamilton vector fields. Additionally in the whole analysis we will neglect boundary terms at asymptotic infinity of Carroll space-time.

Symplectic techniques in the context of field theories, generally applied for covariant field theories, superstring theories and gravity go in the name of covariant phase space \footnote{Technically the space of solutions with the pre-symplectic structure is same as that. However we will omit the term covariant, as there is no manifest covariance in the analysis.} analysis \cite{Julia:2002df, Crnkovic:1986ex, Ashtekar:1990gc, Crnkovic:1986be, Crnkovic:1987tz}. In the Appendix we apply these techniques for Maxwell electrodynamics on an arbitrary Riemann manifold to keep a parallel with the present analysis. There we show that Killing vector fields do give rise to gauge invariant Hamilton functions (which are conserved as well) and establish a homomorphism between the Lie algebra of Killing vectors and the algebra of corresponding Hamilton functions induced by the pre-symplectic structure. This is to exemplify an alternative to the Noether procedure, whose parallel is followed in our system of Carrollian electrodynamics. 

As a first step towards this, we observe that the on-shell variation of the Lagrangian \eqref{carlag} gives the pre-symplectic potential:
\begin{align}
\d L = \p_t \int_{\Sigma} d^dx \, 2\left( \dot{A_i} \d A_i - \p \cdot A \d B \right) =: \p_t \Theta (\delta),
\end{align}
where in the phase space sense $\d$ is to viewed as an arbitrary tangential vector field. 

The pre-symplectic structure, which is a 2-form on phase space, contracted by two arbitrary commuting vector fields $\d_1, \d_2$ is given by:
\begin{align} \label{presymp}
\Omega (\delta_1, \delta_2) = \delta_1 \Theta (\d_2) - (1 \leftrightarrow 2) =  2\int_{\Sigma} d^d x \left(\d_1 \dot{A}_i \d_2 A_i - \p_i \delta_1 A_i \d_2 B - (1 \leftrightarrow 2) \right),
\end{align}
which may be compared with the coordinate representation of differential forms on finite dimensional manifolds. At this point it is easy to verify that the gauge transformations \eqref{cargt1} indeed are degenerate directions of \eqref{presymp}:
\begin{align} \label{gaugetr}
\Omega (\d, \d_{\alpha} ) &= 2 \int d^dx [\d \dot{A}_i \p_i \alpha_2 - \dot{\alpha}_1\d \p \cdot A  - \p_i \dot{\alpha}_2 \d A_i + \p^2 \alpha_2 \d B] \non \\
& =0
\end{align}
provided we use equations of motion \eqref{eleom}, throw away boundary terms at $\p \Sigma$ and identify $\alpha_1$ and $\alpha_2 $ up to space-time constant as required in \eqref{cargt1}. \footnote{Therefore the gauge invariant phase space should be the space of solutions of \eqref{eleom} quotiented by gauge orbits defined via \eqref{cargt1} and the symplectic structure should be a pull back of \eqref{presymp} on that space. However for the practical purpose of defining Hamilton function corresponding to our space-time symmetries, that is not essential as long as we will be working with transformations orthogonal to gauge directions.}

\subsubsection{Hamilton functions and conserved charges}
We are now ready to check whether the infinite dimensional Carrollian conformal algebra \eqref{cca-finite}, \eqref{cca} does have a Hamiltonian action on the phase space. A particular field transformation (or a vector field in the space of solutions) $\d_\star$ is locally Hamiltonian if 
\begin{align} \label{integ}
 \Omega (\d, \d_\star) = \d Q[\star],
\end{align}
ie, an exact variation of a phase space function.
%would now consider the Carrollian kinematical symmetries algebra. This is infinite dimensional because of time translations which are arbitrary spatially dependent or rather super-translations. 
Let us concentrate on the most interesting part, the infinite dimensional Abelian ideal of CCA. They act on our field variables as stated in \eqref{symm_action}:
\bea \label{transcar}
&& \d_f B =  f \dot{B} + \kappa_1 \p_i f A_i \non \\
&& \d_f A_i =  f \dot{A}_i + \kappa_2 \p_i f B
\eea
These transformations indeed generate an Abelian Lie-algebra independent of the values of $\kappa_1, \kappa_2$, in a sense that $[\d_{f_1}, \d_{f_2}] =0$ on any field variable. However on the phase space they define a Hamiltonian flow in the sense of \eqref{integ}, (in other words the right hand side becomes phase space integrable) only if 
\begin{align} \label{kappa_cond}
\kappa_1 =0 , \kappa_2 =1 .
\end{align}
This is important in view of the the present phase space analysis, because this restriction of parameters here was found just by demanding the transformations \eqref{transcar} to be symmetries of the pre-symplectic structure and not even appealing to the symmetry of the action. This is in contrast the previous analysis \cite{Bagchi:2016bcd} where these restrictions emerged from a stricter criterion of being symmetries of equation of motion \eqref{eleom}. Finally, for $\kappa_1 =0 , \kappa_2 =1$ the resulting Hamilton function for arbitrary $f$ is given by:
\bea \label{momentmap}
&&\Omega (\d, \d_f) = \d Q[f] \non \\
&& Q[f] =  \int d^d x \, f (x) \left( \dot{A}_i - \p_i B\right)^2 .
\eea
%It is interesting to note that with these chosen values, the choice of the function as $f=1, x^i$ and $x^2$ respectively make the transformation \eqref{transcar} those of time translation, Carrollian boost and temporal part of the Carrollian special conformal transformation. 

It is easy to see that by the equation of motion, $\frac{d}{dt} Q[f] = 0$. Hence the Carrollian super-translations are symmetries of the system and the Hamilton functions $Q[f]$ are the conserved charges. %The choice $f=1$ gives the total energy of the system. Viewed as a moment map, realization of the Abelian kinematical symmetry algebra can be checked easily via \eqref{momentmap}.

Next we move on to the finite part of CCA and carry on the same analysis for the finite set of generators. Spatial translation and rotation acts on the fields trivially as:
\begin{align}
& \d_p B = p^i \p_i B, \,  \d_p A_j = p^i \p_i A_j, \non \\
& \d_{\omega}B = \omega^{ij}(x_i \p_j- x_j \p_i) B ,\,  \d_{\omega}A_l  = \omega^{ij}\left[(x_i \p_j- x_j \p_i) A_l + \d_{l[i} A_{j]}\right], 
\end{align}
As expected these transformations both generate Hamilton flow:
\bea
\Omega(\d , \d_p) =\d Q[p], \, \Omega(\d , \d_{\omega}) =\d Q[\omega]
\eea
and the Hamilton functions are given by:
\begin{subequations}
\begin{align}
& Q[p] = 2 p^l \int_{\Sigma} d^d x \left( \dot{A}_i \p_l A_i - \p \cdot A \p_l B \right) \\
& Q[\omega] = 2 \omega^{ij} \int_{\Sigma} d^d x \left(\dot{A}_k x_{[i} \p_{j]}A_k + \dot{A}_k \d_ {k[i}A_{j]} - x_{[i} \p_{j]}B \,\p \cdot A  \right)  . 
\end{align}
Using the equations of motion, it can be checked that both the Hamilton functions are conserved:
$$\dfrac{dQ[p]}{dt} =0 = \dfrac{dQ[\omega]}{dt} .$$
\end{subequations}
While checking whether the dilatation transformation corresponds to Hamilton vector fields on phase space we uncover another crucial fact regarding the parameters of the system. As in \eqref{symm_action} we rewrite the corresponding dilatations:
\begin{align*}
& \d_{\Delta} A_ i = t \dot{A}_i + x^j \p_j A_i + \Delta A_i \\
& \d_{\Delta } B = t \dot{B} + x^i \p_i B + \Delta B.
\end{align*}
Plugging this in \eqref{presymp} and using equations of motion \eqref{eleom}, we observe that the expression $\Omega(\d ,\d_{\Delta})$ is phase space integrable only if 
\begin{align} \label{ddel}
\Delta =\frac{d-1}{2}.
\end{align}
This directly implies that the phase space structure is scale invariant in all dimensions provided we scale the fields appropriately obeying \eqref{ddel}. This is similar to the result for the case of free electrodynamics in the relativistic setting. The corresponding Hamilton function is given by:
\begin{align}
& \Omega(\d ,\d_{\Delta}) = \d Q[\Delta] \non \\
& Q[\Delta] = \int_{\Sigma} d^dx [ t \dot{A}^2_i + 2 x^j \dot{A}_i \p_j A_i + t B \p^2 B - 2x^j \p_j B  \p \cdot A + 2\Delta (A_i \dot{A}_i - B \p \cdot A) ]
\end{align}
which also satisfies $\dfrac{dQ[\Delta]}{dt} =0$, making this a conserved charge. The remaining transformation is that due to the spatial part of the special conformal transformation \eqref{spc1}, \eqref{spc2}:
\begin{align} 
\d_{k} B (x,t)=& 2k^i(\Delta x_{i}+ x_{i}t\partial_{t}+x_{i}x^{j}\partial_{j} -\frac{1}{2}x^{j}x_{j}\partial_{i})B  \non \\
 \d_{k}A_{j}(x,t)= & 2k^i(\Delta x_{i} +x_{i}t\partial_{t}+x_{i}x^{l}\partial_{l}- \frac{1}{2}x^{l}x_{l} \partial_{i})A_{j}  \non\\ 
              &  +2 k_j  x^{l} A_{l}-2 k^i x_{j}A_{i} + 2 k_j\, t B \non
\end{align}
imposing the criteria \eqref{kappa_cond} $\kappa_1 =0, \kappa_2 =1$. While going on checking whether the above transformations are Hamiltonian, we further impose \eqref{ddel}. A tedious but straightforward analysis now yields one more condition:
\begin{align} \label{ddel2}
\Delta = d -2
\end{align}
that has to hold for phase space integrability of $\Omega (\d, \d_{k})$. Coupled with \eqref{ddel} we see that this happens for $d=3, \Delta =1$. In other words, Carrollian special conformal transformation generates Hamiltonian flow only in 3 spatial dimensions for (Carrollian) electrodynamics, which is a reconfirmation of the finding made in \cite{Bagchi:2016bcd} at the level of equations of motion. 
\bea
\Omega (\d, \d_{k}) = && \d Q[k] \\
Q[k] = &&  2 k^i \int_{\Sigma} d^3 x \left[2x^j (\dot{A}_i A_j - \dot{A}_j A_i )+ \dot{A}_j \left(2x_i x^l \p_l - x^2 \p_i + t x_i \p_t +2 x_i\right)A_j \right]\non \\
&&  + 2 k^i\int_{\Sigma} d^3 x\left[ t\left(2\dot{A}_i+\p_i B+ x_i \p^2 B\right)- 2A_i  \right]B .
\eea
Again, as expected $Q[k]$ is time preserved and qualifies as a conserved charge.

We would end this section with the important note about gauge transformation. The conserved charges found primarily as Hamilton functions corresponding to CCA transformations are automatically gauge invariant, albeit implicitly since we have not constructed them out of gauge invariant quantities like field strength . The pre-symplectic structure being degenerate along the gauge directions make it obvious. For example consider a Hamilton function $Q[\star]$ corresponding to a transformation $\d_{\star}$ as in \eqref{integ}. Its gauge transformation is given by:
\begin{align}
\d_{\alpha} Q[\star] = \Omega(\d_{\alpha}, \d_{\star}).
\end{align}
But thanks to \eqref{gaugetr} the right hand side above always vanishes, making $Q[\star]$ gauge invariant.
\subsubsection{Algebra of charges}
In the above section we have checked that the actions of the individual basis elements of the infinite dimensional CCA were Hamiltonian. However there is one more check that has to be preformed which will confirm that the Carroll algebra itself has Hamiltonian action \cite{Woodhouse:1992de} on the phase space. For example if $\alpha, \beta, \gamma$ are elements of CCA such that:
\begin{align} \label{alphbet}
[\alpha,\beta] = \gamma ,
\end{align}
where $\gamma$ can be zero. $\d_{ \alpha}$ etc. are the phase space vector fields that act on the fields in a manner as described in \eqref{symm_action}. As shown above these give rise to Hamiltonian flows on the phase space with $Q[\alpha]$ etc. being the corresponding Hamilton functions upto phase space constants. The action of CCA would be said to be Hamiltonian if the following holds:
\begin{align} \label{qalphbet}
\Omega(\d_{\alpha}, \d_{ \beta}) = -Q[\gamma].
\end{align}
This is a statement of linear homomorphism \footnote{Rather it is an anti-homomorphism \cite{woit} due to the apparent sign difference between \eqref{alphbet} and \eqref{qalphbet} which can be attributed to the definition of the moment map \eqref{integ}.} from the Lie algebra CCA to the space of Hamilton functions. 

In physical terms, this means that the kinematical symmetry algebra of the background space-time is realized dynamically at the level of conserved charges (since in this case the concerned Hamilton functions are conserved quantities). Moreover as there are no second class constraints, the charge algebra is equivalent (see \cite{Crnkovic:1987tz} for details) to the Poisson algebra of charges that we encounter in a canonical framework, when the charges are expressed in terms of canonical phase space variables (fields and their conjugate momenta and constraints).

Let us first perform the check for the most unique part of CCA, the infinite dimensional Abelian ideal generated by the super-translations. Corresponding to two arbitrary such elements $M_f$ and $M_g$ for arbitrary spatial functions $f,g$, we have: 
\bea \label{sutraalg}
\Omega (\d_f, \d_g) = 0.
\eea
\textit{This is exact realization of $[M_f, M_g]=0$ without emergence of a central term. Therefore we have found an infinite set of independent conserved charges which all commute among themselves. The phase space vector fields $\d_f$, for all allowed $f$ define an infinite dimensional submanifold in the phase space.}

For the other brackets let us illustrate the one involving spatial translation and rotation generator:
\bea
\Omega(\d_{\omega}, \d_p) &=& -\d_p Q[\omega] \left( = \d_{\omega} Q[p] \right) \non \\
&=& - 2 Q[\tilde{p}] , \, \qquad \mbox{where }\, \tilde{p}^i = \omega^{ij} p_j.
\eea
This is exact realization of the CCA bracket (cf. \eqref{ca-finite}): $ [\omega^{ij}J_{ij}, p^k P_k] = 2 \tilde{p}^i P_i$. In the similar spirit whole of CCA holds at the level of realization to the algebra of charges.
\section{Conclusion and future directions}
In this work we have studied the action for ultra-relativistic electrodynamics and analyzed the conformal Carroll symmetries of it, which to the best of our knowledge is the first for a Carrollia field theory. Previous attempts towards Carrollian invariant actions were for particles, which generally bear the feature of being non-dynamical unless coupled with interactions \cite{Bergshoeff:2014jla}. Interestingly coadjoint orbit treatment of Carrollian photons, treated as particles themselves were carried out in \cite{Duval:2014lpa}. A symplectic approach was taken to find out conserved quantities and it was found that the infinite number of super-translation do not yield non-trivial conserved quantities. This may be compared with the field theory analysis we present where infinite number of conserved quantities were found.

The next step obviously would be to consider interacting theories, bringing in matter interactions and including non-Abelian gauge fields. Finding infinite number of conserved quantities for such systems would make strong progress in the present program. It should be interesting to study if the quantization of such theories bring in anomaly in super-translation part of CCA.

It would be extremely important to construct explicit field theory models with CCA invariane in 3 dimensions, since that would be a BMS${}_4$ invariant theory. This is pertinent since it should qualify as a concrete realization of field theory dual in 4 dimensional asymptotically flat space-time gravity. In addition to that, it will fill a necessary gap since with the present work, we now have an example of a 4 dimensional field theory and the already studied 2 dimensional one \cite{Barnich:2013yka}.

With the infinite number of conserved charges for a local field theory at hand, probably the biggest question should be about integrability. Since there are infinite number of degrees of freedom in the present system, it is not straightforward to analyze even classical integrability of Carrollian electrodynamics. But we can draw inspiration from the 2 dimensional example of a generic BMS${}_3$ invariant theory, where integrability was studied recently \cite{Fuentealba:2017omf} using the similar tools used in constructing KdV hierarchy for 2D CFT.
\section*{Acknowledgements}
The authors gratefully acknowledges discussions with Glenn Barnich, Arjun Bagchi, Amit Ghosh, Avirup Ghosh and Aditya Mehra. RB acknowledges support by DST (India) Inspire award and in part the Belgian Federal Science Policy Office (BELSPO) through the Interuniversity Attraction Pole P7/37 and a research-and-return grant, in part by the “FWO-Vlaanderen” through the project G020714N, and by the Vrije Universiteit Brussel through the Strategic Research Program “High-Energy Physics”. RB also thanks the hospitality of IIT Kanpur during preparation of the work. 
\begin{appendix}
\section{APPENDIX: Pre-symplectic geometry of Maxwell theory on Riemann manifold}
Consider action of Maxwell electrodynamics on an $n$-dimensional arbitrary (pseudo)-Riemann manifold $(M,g)$, constructed by $U(1)$ gauge fields $A$ which are space-time 1-forms:
\begin{eqnarray}
S = \int_{M}L=  k \int_{M} dA \wedge \star dA
\end{eqnarray}
where $k$ is an real number, unimportant for our purpose and $L$ is the Lagrangian $n$-form.

Let $X$ be a Killing vector field of $(M,g)$. Diffeomorphism (in this case isometric) generated by $X$ act infinitesimally on fields as $\d_X A = \pounds_X A$. Since $X$ is Killing, Lie-derivative $ \pounds_X$ should commute with the Hodge star when acting on differential forms. We use this property throughout this section.
\subsection{Noether procedure}
Off-shell variation of the Lagrangian for a Killing vector $X$:
\begin{align} \label{maxwell_lagrangian}
&\d_X L = d \alpha[X] \non \\
& \alpha[X] = i_X L.
\end{align}
This is the statement of $\d_X$ being a symmetry transformation. The pre-symplectic potential found by first variation of the Lagrangian on-shell is:
\begin{align} \label{presymplectic_potential}
\d L = d\left(\Theta (\delta) \right) = d \left( 2 k \d A \wedge \star dA \right)
\end{align}
We should keep in mind, from the form of the pre-symplectic potential, that the variational principle to hold true from the action corresponding to \eqref{maxwell_lagrangian} one must impose the pull back of the 3-form $\d A \wedge \star dA$ to vanish on the boundaries of space-time which are not time-like. Otherwise one could add boundary term to the action so as to make the variation well defined for all field configurations. Any phase-space (space of solutions of equations of motion $d\star d A =0$) variation $\delta$ will be said to be allowed for the above to happen. For example in the case of asymptotically flat space-time like Minkowski, the conformal boundary is composed of the null-infinities $ \mathcal{I}^+ \cup \mathcal{I}^-$ and their intersection is the spatial infinity $i^0$.

Comparing \eqref{maxwell_lagrangian} and \eqref{presymplectic_potential} we see that on-shell, the space-time $n-1$-form:
\begin{align}
J[X] &= \Theta (\d_X) - \alpha [X] \, \, ~~~~~ \, \mbox{on-shell} \non \\
& = k \left( i_X  dA \wedge \star dA - dA \wedge i_X (\star dA) \right) +2 k \, d \left( i_X A \wedge \star dA \right) 
\end{align}
is a closed form and qualifies as the Noether current corresponding to the symmetry transformation $\d_X$. The corresponding conserved Noether charge is the integrated current (pulled back) to a spatial surface $\Sigma$:
\begin{align} \label{noether_maxwell}
Q_X = k\int_{\Sigma} i_X  dA \wedge \star dA - dA \wedge i_X (\star dA) + 2 k \int _{\d \Sigma} (i_X A ) \star dA 
\end{align}
\subsection{Hamilton vector field analysis}
The pre-symplectic structure is found by second variation of the Lagrangian on top of \eqref{presymplectic_potential} and integrating over a space-like $n-1$ dimensional surface:
\begin{eqnarray} \label{maxsymp}
\Omega(\delta_1, \delta_2) = 2k \int_{\Sigma} \delta_1 A \wedge \star d \delta_2 A - (1 \leftrightarrow 2).
\end{eqnarray}
 Using this \footnote{Discussion with Avirup Ghosh for this derivation is gratefully acknowledged.}, we get:
\begin{eqnarray} \label{charge}
\Omega (\delta, \delta_X) %&=& \int_{\Sigma} \delta A \wedge \pounds_X \star dA - \pounds_X A \wedge \star d\delta A \nonumber \\
%&=& \int_\Sigma d(\delta A) \wedge i_X(\star dA) - i_X (dA) \wedge \star d \delta A \nonumber \\
%&=&\delta \left( \int_\Sigma dA \wedge i_X (\star dA) - \frac{1}{2} i_X \left( dA \wedge \star dA \right) \right) \non \\
&=&  - k\,\delta \int_\Sigma i_X  dA \wedge \star dA - dA \wedge i_X (\star dA)   \non\\
&&- 2k \d \int_{\p \Sigma} (i_X A ) \star dA + 2k \int_{\p \Sigma} i_X \left(\d A \wedge \star d A \right)\non\\
&=&  \delta \tilde{Q}_X + 2k \int_{\p \Sigma} i_X \left(\d A \wedge \star d A \right)
\end{eqnarray}
%In the second line we have used the euations of motion for $A$ and $\delta A$. At the last step we have used $dA \wedge \star d \delta A = d \delta A \wedge \star d A$. Moreover we have dropped boundary terms at different steps. The total boundary term that has been dropped is:
%\bea
%\delta \int_{\p \Sigma} A \wedge i_X \star dA + \int_{\p \Sigma} i_X \left( A \wedge \star d \d A \right) 
%\eea
%The important point to note is that the boundary terms appearing in \eqref{charge} do vanish for appropriate fall-off condition on $A$. For example, in Minowski space, $\p \Sigma$ is actually the spatial infinity $i^{0}$. $\frac{1}{r^{1+\epsilon}}$ for $\epsilon > 0$ fall-off of the gauge field components ensure vanishing of the boundary terms as for all Poincare generators $X$. In that case, trivially the transformations $\d_X$ are Hamiltonian. However, irrespective of fall-off conditions, if $X$ on $i^0$ is a vector field parallel to it, then the last term in \eqref{charge} vanishes, which guarantees:
It is easy to argue that the second term in the right hand side of \eqref{charge} should vanish from variational boundary condition \footnote{However, irrespective of boundary conditions, if $X$ on $i^0$ is a vector field parallel to it, then the last term in \eqref{charge} vanishes.} argument presented above. This guarantees:
\bea
\Omega (\delta, \delta_X) =  \delta \tilde{Q}_X,
\eea 
ie $\d_X$ generates locally Hamiltonian flow on phase space and the corresponding Hamilton functions $\tilde{Q}_X$ is same as the conserved Noether charge $Q_X$ as in \eqref{noether_maxwell} up to a sign. The sign difference reflects our convention of defining Hamilton functions. 

For now, if we consider the boundary term in \eqref{noether_maxwell} to vanish, this can be seen to give the usual charges corresponding to the gauge invariant form of Electromagnetic energy-momentum tensor, for $X = a^{\mu} \p_{\mu}$ ($a^{\mu}$ are space-time constant) in Minkowski background. 

Now let us consider $Y$ to be another Killing vector field of $M,g$. Hence the vector $Z = [X,Y] = \pounds_ X Y$ is also Killing. Disregarding the boundary terms (considering appropriate fall-offs) and using multiple manipulations with Hodge star and differential forms, we recover the desired charge algebra:
\bea
\tilde{Q}_Z = \tilde{Q}_{[X, Y]}&=& - 2k\int_{\Sigma} \lx A \wedge \star d \pounds_Y A - \pounds_Y A \wedge \star d \pounds_X A \nonumber \\
 &=& -\Omega(\delta_X, \delta_Y)
\eea
which is an exact anti-homomorphsim (due to the sign which is a result of our choice of Hamilton functions) of the kinematical algebra: $Z = [X,Y]$
\subsection{The Sky group}
Now consider a vector field in the field space that induces $U(1)$ gauge transformations:
$$ \delta _{ \lambda} A = d \lambda .$$
From the Noether procedure we don't find any non-trivial conserved charge corresponding to this, as expected. We may now wish to see if this transformation generates a Hamiltonian flow in the phase space. To this end, plugging the above transformation in the pre-symplectic structure yields:
\begin{eqnarray}\label{maxg}
\Omega ( \delta, \delta _{\lambda})= -2k \int _{ \partial \Sigma} \lambda \ast d( \delta A)
\end{eqnarray}
For the asymptotically flat cases, $\partial \Sigma$ will be taken to be $i^0$. For gauge transformations $\lambda$ which are state-independent, the $\d$ in the right hand side of \eqref{maxg} can be pulled out of the integral, proving gauge transformations to be Hamiltonian flows corresponding to Hamilton functions:
\begin{align} \label{gauge_charge}
Q_{\lambda} = 2k \int _{ \partial \Sigma} \lambda \ast d A;
\end{align}
 but only with support from $i^0$ (unlike the ones corresponding to global symmetries). Note that since there is no contribution from the bulk, these would simply vanish for strict fall of conditions at the boundary making the phase-space vector field $\d_{\lambda}$ a degenerate direction for \eqref{maxsymp}. 
 
Since $\d_{\lambda}$ were invariance of the action, the corresponding Hamilton functions are also conserved and hence these are conserved charges. The pre-symplectic structure also induces a Lie-algebra on the vector space of charges:
\begin{align}
\Omega (\d_{\lambda_1}, \d_{\lambda_2}) = 0.
\end{align}
This implies that this infinite dimensional Lie algebra is Abelian. The corresponding Lie-group is the Sky group \cite{Balachandran:2014hra} analogous to the gravitational Spi group at $i^0$.
%\subsection{Gauge and boundary effects}
%Gauge variation $\d_{ \lambda} = d \lambda$ is generally a degenerate direction for the pre-symplectic structure in the covariant phase space. But we get physical large $U(1)$ gauge transformation at boundary, give physical charges for non-trivial boundary conditions giving rise to physically meaningful charges:
%\bea \label{gauge_charge}
%\Omega(\d, \d_{\lambda}) = -2k \d \int_{\p \Sigma} \lambda \star dA.
%\eea
%These charges generate the Abelian, infinite dimensional Sky group at $i^0$.

From the last section's discussion we gather that if $X$ is a space-time Killing vector $\d_X$ gives conserved Hamilton function. The canonical bracket between such a charge with the boundary physical charge \eqref{gauge_charge} is given by:
\bea
\Omega (\d_X, \d_{\lambda}) = 2k \int_{\p \Sigma} d \lambda \wedge i_X \star dA
\eea 
This means that large gauge transformations, which are not constant on $\p \Sigma$ do actually break  Lorentz symmetry in Minkowski space-time.
\end{appendix}

\end{document}